\documentclass[pageno]{style/jpaper}

\usepackage{cite}
\usepackage{booktabs}
\usepackage[normalem]{ulem}
\usepackage{graphicx}
\usepackage{amsmath}
\usepackage{cleveref}
\usepackage{enumitem}
\usepackage{textcomp}
\usepackage{inconsolata}
\usepackage{float}
\usepackage[ruled,linesnumbered]{algorithm2e}

\usepackage{listings, xcolor}

\definecolor{verylightgray}{rgb}{.97,.97,.97}

\lstdefinelanguage{Solidity}{
	keywords=[1]{anonymous, assembly, assert, balance, break, call, callcode, case, catch, class, constant, continue, constructor, contract, debugger, default, delegatecall, delete, do, else, emit, event, experimental, export, external, false, finally, for, function, gas, if, implements, import, in, indexed, instanceof, interface, internal, is, length, library, log0, log1, log2, log3, log4, memory, modifier, new, payable, pragma, private, protected, public, pure, push, require, return, returns, revert, selfdestruct, send, solidity, storage, struct, suicide, super, switch, then, this, throw, transfer, true, try, typeof, using, value, view, while, with, addmod, ecrecover, keccak256, mulmod, ripemd160, sha256, sha3}, 
	keywordstyle=[1]\color{blue}\bfseries,
	keywords=[2]{address, bool, byte, bytes, bytes1, bytes2, bytes3, bytes4, bytes5, bytes6, bytes7, bytes8, bytes9, bytes10, bytes11, bytes12, bytes13, bytes14, bytes15, bytes16, bytes17, bytes18, bytes19, bytes20, bytes21, bytes22, bytes23, bytes24, bytes25, bytes26, bytes27, bytes28, bytes29, bytes30, bytes31, bytes32, enum, int, int8, int16, int24, int32, int40, int48, int56, int64, int72, int80, int88, int96, int104, int112, int120, int128, int136, int144, int152, int160, int168, int176, int184, int192, int200, int208, int216, int224, int232, int240, int248, int256, mapping, string, uint, uint8, uint16, uint24, uint32, uint40, uint48, uint56, uint64, uint72, uint80, uint88, uint96, uint104, uint112, uint120, uint128, uint136, uint144, uint152, uint160, uint168, uint176, uint184, uint192, uint200, uint208, uint216, uint224, uint232, uint240, uint248, uint256, var, void, ether, finney, szabo, wei, days, hours, minutes, seconds, weeks, years},	
	keywordstyle=[2]\color{teal},
	keywords=[3]{block, blockhash, coinbase, difficulty, gaslimit, number, timestamp, msg, data, gas, sender, sig, value, now, tx, gasprice, origin},	
	keywordstyle=[3]\color{violet}\bfseries,
	identifierstyle=\color{black},
	sensitive=false,
	comment=[l]{//},
	morecomment=[s]{/*}{*/},
	commentstyle=\color{gray}\ttfamily,
	stringstyle=\color{red}\ttfamily,
	morestring=[b]',
	morestring=[b]"
}

\newcommand{\name}{Olive}
\crefname{algocf}{Algorithm}{Algorithms}
\newcommand{\code}[1]{\texttt{{\small \detokenize{#1}}}}

\begin{document}

\title{\textbf{Operation-level Concurrent Transaction Execution for Blockchains}}
\author{
\fontsize{13}{14}\selectfont Haoran Lin\\ 
\fontsize{11}{14}\selectfont \nolinkurl{haoran\_lin@zju.edu.cn} \\ 
\fontsize{11}{14}\selectfont Zhejiang University
\and 
\fontsize{13}{14}\selectfont Yajin Zhou\\ 
\fontsize{11}{14}\selectfont \nolinkurl{yajin@blocksec.com} \\ 
\fontsize{11}{14}\selectfont Zhejiang University \\ 
\fontsize{11}{14}\selectfont BlockSec \\
\and
\fontsize{13}{14}\selectfont Lei Wu\\ 
\fontsize{11}{14}\selectfont \nolinkurl{lwu@blocksec.com} \\
\fontsize{11}{14}\selectfont Zhejiang University \\ 
\fontsize{11}{14}\selectfont BlockSec 
}
\date{}

\maketitle

\thispagestyle{empty}

\begin{abstract}

Despite the success in various scenarios, blockchain systems, especially EVM-compatible ones that serially execute transactions, still face the significant challenge of limited throughput.
Concurrent transaction execution is a promising technique to accelerate transaction processing and increase the overall throughput.
Existing concurrency control algorithms, however, fail to obtain enough speedups in real-world blockchains due to the high-contention workloads.

In this paper, we propose a novel operation-level concurrency control algorithm designed for blockchains.
The core idea behind our algorithm is that only operations depending on conflicts should be executed serially, while all other conflict-free operations can be executed concurrently.
Therefore, in contrast to the traditional approaches, which block or abort the entire transaction when encountering conflicts, our algorithm introduces a redo phase to resolve conflicts at the operation level by re-executing conflicting operations only.
We also develop a set of data dependency tracking mechanisms to achieve precise identification and speedy re-execution for conflicting operations.
We implement an open-source prototype based on Go Ethereum and evaluate it using real-world Ethereum blocks.
The evaluation results show that our algorithm achieves an average speedup of 4.28$\times$.
If combined with state prefetching techniques, our approach can further accelerate the transaction execution by 7.11$\times$.
\end{abstract}

\section{Introduction}
Following the growing prosperity of cryptocurrencies~\cite{nakamoto2008bitcoin,wood2014ethereum}, blockchain technologies have gained increasing attention.
The introduction of programable smart contracts~\cite{smart-contract} has further expanded the applications of blockchains beyond cryptocurrencies, e.g., decentralized finance.
But the limited throughput remains one of the most critical issues of blockchains, hindering their wider adoption.

The throughput of blockchains can be increased by either shortening block time or enlarging block size.
Block time (i.e., the average time to generate a new block) is typically determined by the consensus protocol.
In recent years, researchers have proposed many consensus protocols capable of generating a block in a short time~\cite{sompolinsky2018phantom,conflux,sompolinsky2015secure,lewenberg2015inclusive}.
Ethereum~\cite{wood2014ethereum}, one of the most popular blockchain platforms, has already switched to the efficient Proof-of-Stake (PoS) consensus~\cite{king2012ppcoin} from the slow Proof-of-Work (PoW) consensus~\cite{nakamoto2008bitcoin}.
With these advances in consensus protocols, the block time is no longer the throughput bottleneck.
On the other hand, block size is determined by the number of transactions executed within the block time.
If the transaction execution speed remains the same, a shorter block time eventually leads to a smaller block size, which is unfavorable for the overall throughput.
Therefore, optimizing consensus protocols alone does not necessarily improve the throughput.
But if the transaction execution bottleneck is alleviated, blockchains can be configured with both short block time and large block size, thus increasing throughput.

\noindent\textbf{Concurrent Execution in Blockchains.}
One of the most promising techniques to accelerate the execution layer of blockchains is \emph{concurrent execution}.
It harnesses the parallelism of commodity hardware, which is untapped under the serial execution model, to speed up transaction execution.
Recent works~\cite{dickerson,optsmart,cc-permissoned,garamvolgyi2022utilizing} have attempted to adopt traditional concurrency control algorithms to blockchains.
But they fail to obtain enough speedups in real-world blockchain workloads due to the \emph{hot spot problem} (that causes a large number of conflicts).

Blockchain workloads feature highly skewed data accesses, making hot spots common.
For example, in Ethereum workloads, 62\% storage operations access only 0.1\% of storage slots.
Moreover, most storage operations in blockchains follow the read-modify-write (RMW) pattern.
Multiple concurrent RMW operations accessing data hot spots bring about severe data contention and lead to significant performance degradation for traditional concurrency control algorithms.
Specifically, pessimistic algorithms, such as two-phase locking (2PL)~\cite{2pl}, only obtain low concurrency due to the lock contention on hot spots; while optimistic algorithms, such as optimistic concurrency control (OCC)~\cite{occ}, encounter frequent aborts due to data conflicts on hot spots.

The root causes behind the poor performance of these traditional algorithms are their \emph{transaction-level} conflict-handling strategies.
They view transactions as indivisible and treat internal operations as "black boxes."
When encountering a conflict, they have to block or abort the \emph{entire} transaction, regardless of how many operations are affected by the conflict.
Such coarse-grained parallelizing strategies are inefficient, especially in high-contention blockchain workloads.

\begin{figure}[t]
    \centering
    \includegraphics[width=0.28\textwidth]{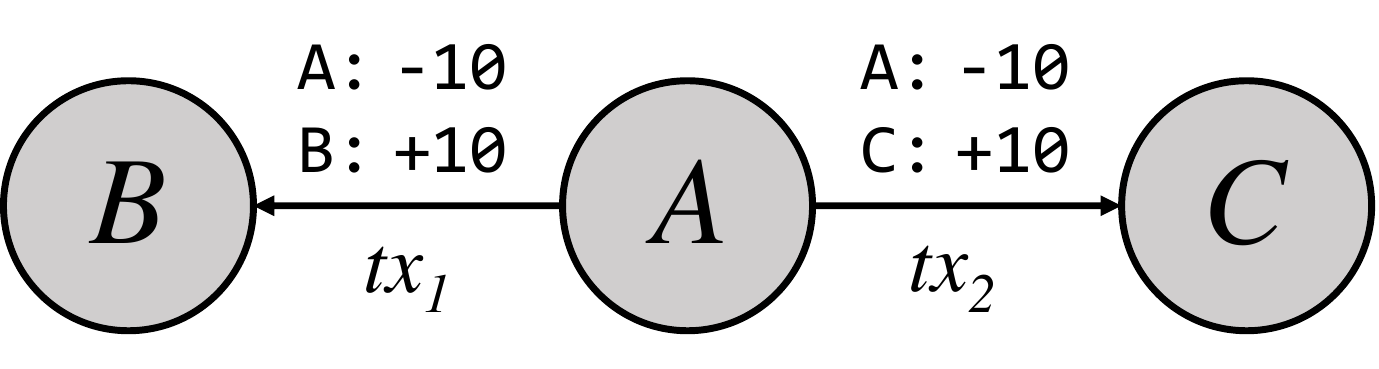}
    \caption{Token transfer example}
    \label{fig:transfer}
\end{figure}

\noindent\textbf{Key Insights.}
To motivate a new conflict-handling strategy, let us consider a typical token transfer example shown in \Cref{fig:transfer}.
Transaction $tx_1$ manifests a transfer of ten tokens from user $A$ to $B$, and transaction $tx_2$ manifests a transfer of ten tokens from user $A$ to $C$.
These two transactions conflict because they both read, modify and write $A$'s balance.
But it is worth noting that the data conflict on $A$'s balance does not influence the balances of $B$ and $C$.
They are increased by ten tokens regardless of the conflict, as long as $A$ has enough tokens to complete both $tx_1$ and $tx_2$.
In other words, the RMW operations on the balances of $B$ and $C$ are conflict-free and thus can be executed concurrently without blocking or aborting.
To achieve this, a possible optimistic execution schedule can be:
(i) execute $tx_1$ and $tx_2$ concurrently and speculatively just as OCC,
(ii) validate and commit $tx_1$,
(iii) detect that $tx_2$ misses $tx_1$'s update on $A$'s balance,
and (iv) re-execute the RMW operations on $A$'s balance in $tx_2$ and then commit $tx_2$.
In this way, we resolve data conflicts at the \emph{operation level}.
That is, instead of blocking or aborting the entire transaction, we only re-execute operations that depend on the conflicting data (i.e., the update on $A$'s balance in $tx_2$ in this example).

\noindent\textbf{Our Approach.}
With the insights above, we propose a novel \emph{operation-level} concurrent control algorithm capable of exploiting parallelism even in high-contention blockchain workloads.
The central idea behind our algorithm is to resolve data conflicts at the operation level: only operations depending on the conflicting data need to be executed serially, while all other conflict-free operations can be executed concurrently without blocking or aborting.

Specifically, we use a variant of optimistic concurrency control (OCC).
Like OCC, our algorithm first executes transactions concurrently and speculatively while tracking storage slots they read and write in the thread-local memory.
Then our algorithm validates and commits these transactions one by one.
If a transaction's read set is consistent with the committed storage, our algorithm commits the transaction and installs all values in its write set.
Otherwise (i.e., conflicts are detected), rather than aborting the transaction as OCC, our algorithm introduces a novel \emph{redo phase} to identify and re-execute all operations influenced by the conflicts.
In this way, our algorithm resolves data conflicts by re-executing only several operations, which largely mitigates the performance degradation caused by data contention.

Moreover, to achieve conflicting operation identification and re-execution in the redo phase, we propose a technique called \emph{SSA (static single assignment) operation log}.
Unlike stack-based EVM instruction traces, our SSA operation logs not only record the inputs and outputs of the executed operations but also maintain the dependencies (i.e., use-definition and definition-use chains) among operations.
With the help of the SSA operation log, we can identify all conflicting operations by traversing the definition-use chains, retrieve their inputs from the definition operations, and finally re-execute them.

We have implemented a prototype system, \name{}, based on the Go Ethereum~\cite{geth}, and we will release the source code to engage the community\footnote{\url{http://github.com/XXXX/XXXX}}.
We conduct a comprehensive evaluation of our prototype system.
The evaluation results show that \name{} achieves an average speedup of 4.28$\times$ on real-world Ethereum blocks.
Furthermore, if combined with state prefetching techniques, the speedup can reach 7.11$\times$.

\noindent\textbf{Our Contribution.}
In summary, this paper makes the following contributions:
\begin{itemize}
    \item \textbf{Novel Strategy.} We introduce a novel operation-level conflict-handling strategy as the enabling mechanism for parallelizing transaction operations even in high-contention blockchain workloads.
    \item \textbf{New Techniques.} We propose a new SSA operation log for conflicting operation identification and re-execution and an effective data-flow tracking algorithm for the SSA operation log generation.
    \item \textbf{Efficient Prototype.} We implement a prototype system, \name{}, which can process real-world Ethereum transactions in parallel.
    To the best of our knowledge, \name{} is the first implementation of Ethereum Virtual Machine (EVM) that supports operation-level concurrent transaction execution.
    \item \textbf{Thorough Evaluation.} We evaluate \name{} using real-world Ethereum blocks. 
    The results demonstrate that \name{} accelerates the execution layer of Ethereum effectively and creates a real opportunity to increase the overall throughput.
\end{itemize}
\section{Background}
\subsection{Blockchain}
A blockchain is a replicated state machine shared across computers on a peer-to-peer network.
It consists of an ordered list of \emph{blocks} linked via cryptographic hash pointers.
Each block further comprises a sequence of \emph{transactions}, which either transfer cryptocurrency or invoke \emph{smart contract}~\cite{wood2014ethereum} code to modify the blockchain state.
The design of a blockchain system typically involves two independent layers: the consensus layer that ensures mutually distrustful nodes agree on the same block history and the state layer that defines the state structure and transition rules.

\begin{figure}[htbp]
    \centering
    \includegraphics[width=0.45\textwidth]{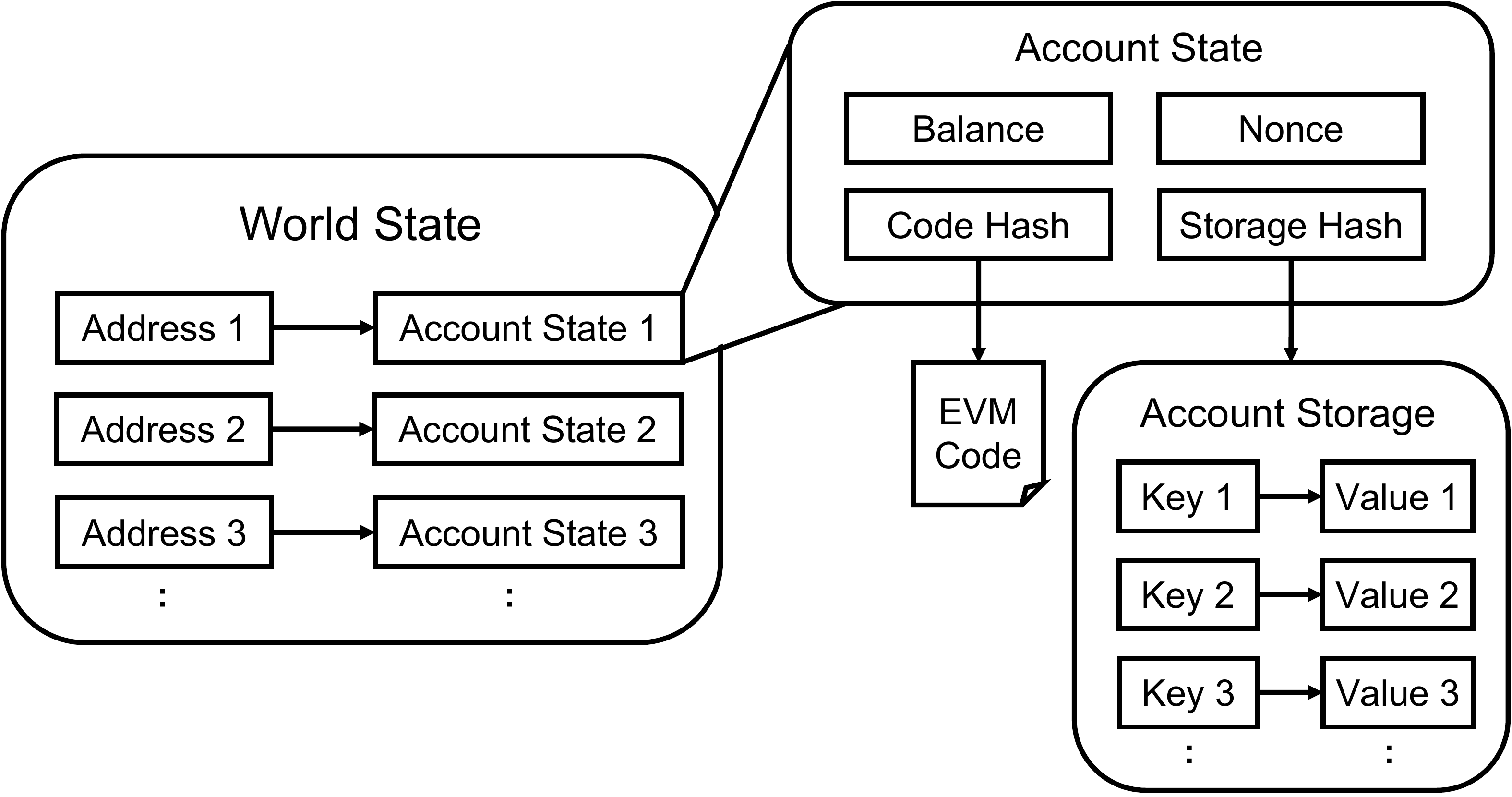}
    \caption{Ethereum world state}
    \label{fig:world_state}
\end{figure}

\noindent\textbf{Ethereum's State.}
Ethereum maintains its state as so-called the \emph{world state}~\cite{wood2014ethereum}.
As depicted in \Cref{fig:world_state}, it is a mapping between the account address and the account state.
An Ethereum account state consists of four fields: 
(1) balance, the number of Ether (the native cryptocurrency of Ethereum) owned by this account, 
(2) nonce, the number of transactions sent by this account, 
(3) code hash, the hash pointer of the smart contract code of this account,
and (4) storage hash, the hash pointer of the root node of a Merkle Patricia tree~\cite{mpt} that encodes the key-value storage (the mapping between 256-bit integers) of this account.

The Ethereum's world state transition rules are defined by the \emph{Ethereum Virtual Machine (EVM)}~\cite{wood2014ethereum}, a quasi-Turing-complete machine.
EVM is a 256-bit word stack machine with volatile byte-addressable memory and persistent key-value storage.
The EVM code can access the stack via \code{PUSH}, \code{POP}, \code{SWAP}, \code{DUP}, etc., the memory via \code{MLOAD}, \code{MSTORE}, \code{MSTORE8}, etc., and the storage via \code{SLOAD} and \code{SSTORE}.
When a transaction invokes a smart contract, the EVM executes the contract code and manipulates the world state.
Each EVM execution step incurs a cost denominated in gas, and the transaction sender pays the total amount of gas consumed by the transaction.

\noindent\textbf{Throughput Bottleneck.}
The throughput of a blockchain is determined by two factors: the \emph{block time} and the \emph{block size}.
The block time is typically set as a constant at the consensus layer and has been considered the throughput bottleneck.
For example, Proof-of-Work (PoW) consensus~\cite{nakamoto2008bitcoin}, introduced by Bitcoin, must keep a slow block generation rate (about 10 minutes a block) to avoid the generation of concurrent blocks.
As a result, Bitcoin processes only about seven transactions per second.
Nevertheless, in recent years, many consensus protocols which achieve fast block generation rates have been proposed and adopted in both industry and academia~\cite{sompolinsky2018phantom,conflux,sompolinsky2015secure,lewenberg2015inclusive}.
For example, Ethereum has already switched to the efficient Proof-of-Stake (PoS) consensus~\cite{king2012ppcoin}, which enables short block time.~\footnote{For compatibility, the block time for Ethereum using PoS is still around 12 seconds at the current stage.}
Moreover, Conflux, a permissionless blockchain, can generate a block in less than 0.1 seconds~\cite{conflux}.
With these advances, the consensus layer is no longer the performance bottleneck.

To further improve throughput, researchers then attempt to increase the block size.
However, the block size cannot be arbitrarily large.
Otherwise, the less performant full nodes would not catch up with the blockchain due to the \emph{storage space} and \emph{execution speed} requirements~\cite{block}.
Sharding is a promising approach to address the space issue~\cite{cosplit,ELASTICO,elrondhighly,monoxide,OmniLedger}.
It splits the blockchain state into multiple shards so that every node holds only a portion of the overall state.
Nevertheless, sharding does not solve the speed issue due to the \emph{hot spot problem}.
Most transactions in a blockchain invoke only several highly-popular contracts and thus are likely to be processed by only several hot shards.
That is, nodes in hot shards still have to execute a large number of transactions, which severely restricts the block size increment.
Therefore, \emph{transaction execution acceleration is a crucial step toward high-throughput blockchains}.

\noindent\textbf{Limitation of Serial Execution.}
Most blockchain systems execute transactions within a block serially.
Such serial transaction execution models are easy to reason about but fail to exploit the internal parallelism of modern commodity hardware (e.g., multi-core processors and SSDs).
This motivates us to introduce \emph{concurrent transaction execution}.


\subsection{Concurrency Control Algorithm}
\noindent\textbf{Concurrency Control in DBMSs.}
Traditional DBMSs give transactions the illusion that their execution is isolated from other transactions, i.e., they cannot observe the effects of concurrent transactions.
Researchers have proposed various \emph{concurrency control algorithms} to execute transactions concurrently while maintaining the illusion of isolation.

The core of a concurrency control algorithm is its strategy for handling conflicts.
A \emph{conflict} between two operations of different transactions occurs if they access the same object and at least one is a write.
We classify existing concurrency control algorithms into two categories based on their conflict-handling strategies:

\begin{itemize}
    \item \emph{Pessimistic algorithms}, which use locks to avoid conflicts.
    They do not allow a transaction to access an object until the corresponding lock is acquired.
    The most frequently used pessimistic scheme is the two-phase locking (2PL)~\cite{2pl}.
    As its name suggests, 2PL divides transaction execution into two phases: 
    (1) a growing phase where transactions only acquire locks on demand (i.e., no lock releasing), and (2) a shrinking phase where transactions only release locks (i.e., no lock acquiring).
    \item \emph{Optimistic algorithms}, which assume conflicts between transactions are rare. 
    They execute transactions concurrently without blocking any operation but abort and restart conflicting transactions once conflicts are detected.
    \emph{Optimistic concurrency control (OCC)}~\cite{occ} is a typical optimistic algorithm.
    The execution of an OCC transaction consists of three phases:
    (1) a read phase, where the DBMS executes the transaction in its private workspace and tracks its read and write sets,
    (2) a validation phase, where the DBMS examines whether the transaction's read set overlaps with other concurrent transactions' write sets,
    and (3) a write phase, where the DBMS installs the transaction's write set if the validation succeeds or otherwise aborts and restarts the transaction.
\end{itemize}

\noindent\textbf{Concurrency Control in Blockchains.}
Unlike DBMSs, blockchains impose an additional restriction on transaction execution --- transactions must commit in the order specified in the block.
Researchers have revised these traditional concurrency control algorithms so that they can be employed in blockchains~\cite{dickerson,optsmart,cc-permissoned,garamvolgyi2022utilizing}:
\begin{itemize}
    \item For pessimistic algorithms like 2PL, every transaction must hold all acquired locks until it commits.
    Moreover, when a transaction $T_1$ attempts to acquire a lock held by another transaction $T_2$, $T_1$ must wait until $T_2$ commits if $T_1$ should commit after $T_2$; otherwise, $T_1$ acquires the lock immediately by aborting and restarting $T_2$.
    \item For optimistic algorithms like OCC, a transaction can enter the validation phase only if all previous transactions commit.
\end{itemize}
\section{Motivation}
\subsection{The Hot Spot Problem}
\begin{figure}
    \subfloat[Hot contracts]{
        \includegraphics[width=0.45\textwidth]{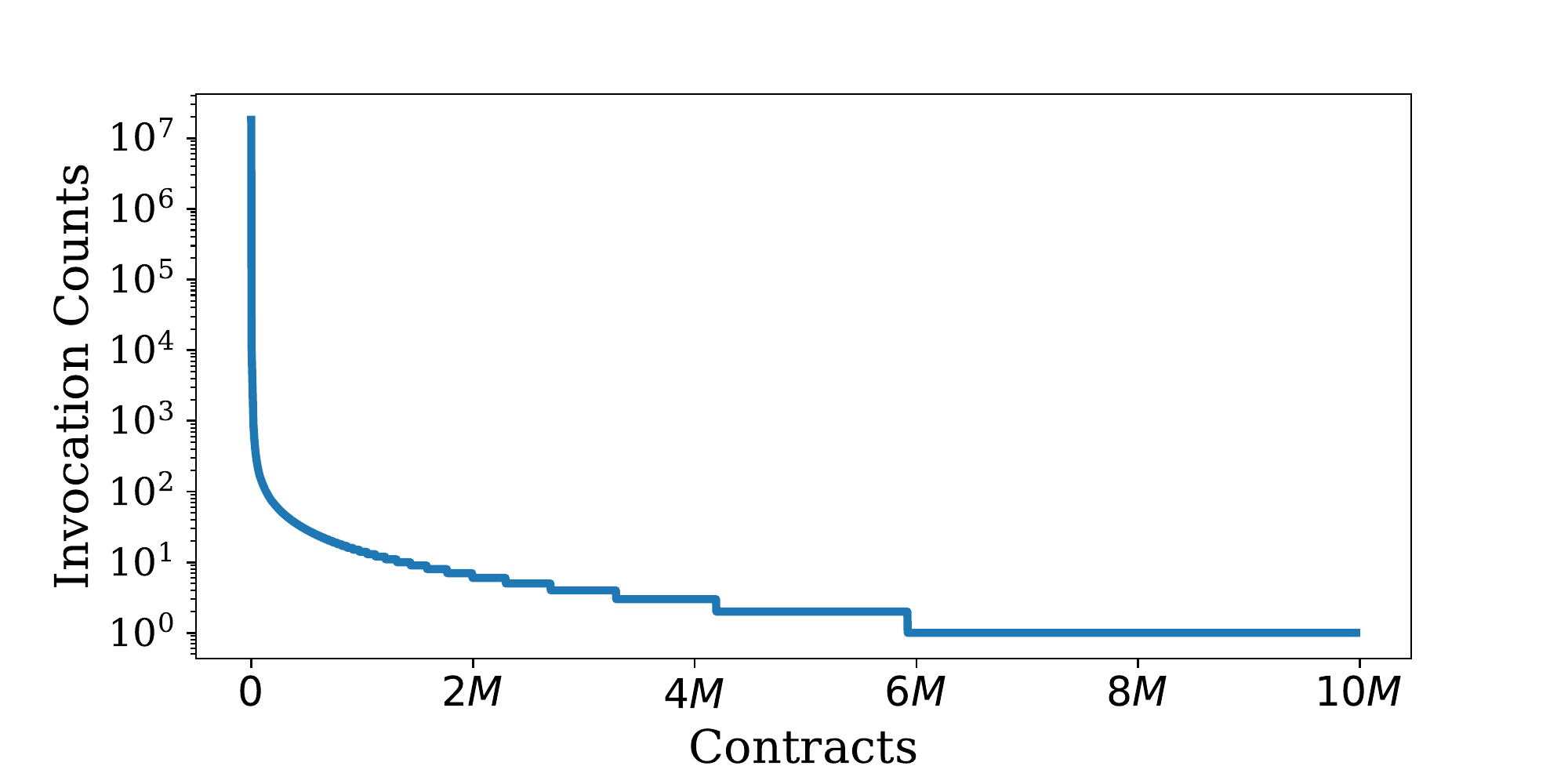}
        \label{fig:hot_contract}
    }
    
    
    \subfloat[Hot storage slots]{
    \includegraphics[width=0.45\textwidth]{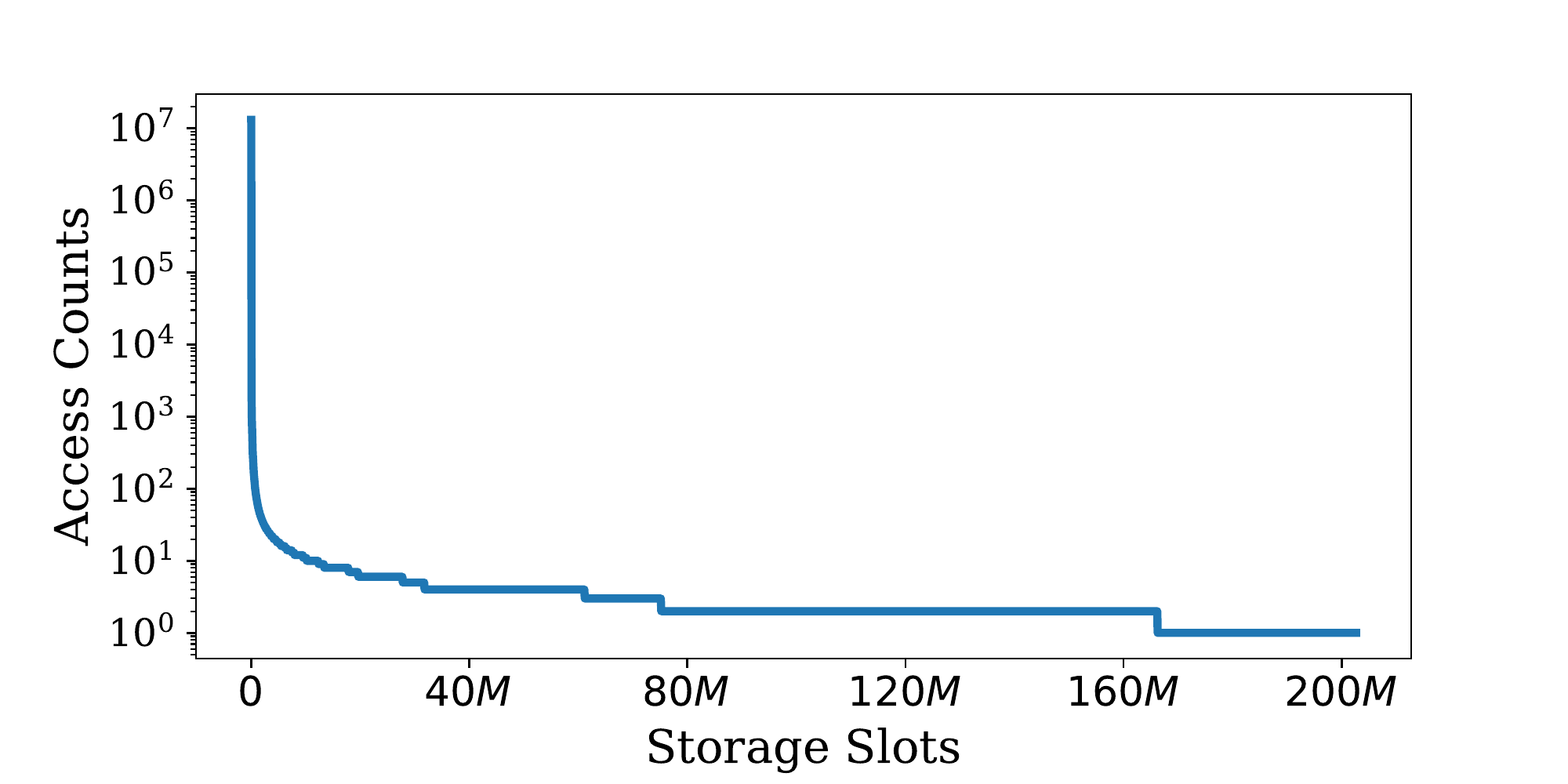}
    \label{fig:hot_slot}
    }
    
    \caption{The hot spot phenomenons for (a) contracts and (b) storage slots.
    The contract and slot indices (X-axises) are sorted in the descending order in terms of their invocation and access counts.
    The invocation and access counts (Y-axises) are plotted on a logarithmic scale.
    }
    \label{fig:hotspot}
\end{figure}

The most prominent characteristic of blockchain workloads is the \emph{hot spot problem}.
Data accesses in blockchains are highly skewed: a few contracts or storage slots are invoked or accessed far more often than others.
For example, some popular contracts, such as CryptoKitties and crowdfunding contracts, had congested the Ethereum network for noticeable periods.
To quantitatively illustrate the phenomenon, we collect the invocation and access counts of each distinguished contract and storage slot on Ethereum from Jan-01-2022 to Jul-01-2022.
\Cref{fig:hotspot} shows the results: among 10 million contracts, 0.1\% of contracts account for 76\% of invocation counts; among 200 million storage slots, 0.1\% of storage slots account for 62\% of storage access counts.
Moreover, we find that the top ten contracts contribute to about 25\% of contract invocations, and nine of them are ERC20 contracts~\cite{erc20}, the most popular fungible token standard on Ethereum.
These results demonstrate the severe data contention of blockchain workloads.

The hot spot problem degrades the performance of traditional concurrency control algorithms significantly.
For blockchain-versioned pessimistic algorithms, a transaction may be blocked for a long time or even aborted by other higher-priority transactions due to lock contention.
For the optimistic algorithms, severe data conflicts lead to numerous aborted transactions.
Therefore, \emph{how to design an efficient concurrency control algorithm to deal with the hot spot problem in blockchain workloads} remains an open problem.


\subsection{Toward Parallelizing an ERC20 Contract}\label{sec:erc20}
\begin{figure}
\begin{lstlisting}[language=Solidity]
contract ERC20 {
  /* persistent states */
  mapping(address => uint256) balances;
  mapping(address => mapping(address => uint256)) allowances;
  
  /* functions */
  function transferFrom(from, to, amount) {
    _useAllowance(from, _msgSender(), amount);
    _transfer(from, to, amount);
  }
  function _transfer(from, to, amount) {
    ...
    require(balances[from] >= amount);(*@\label{code:balance_check}@*)
    balances[from] -= amount;(*@\label{code:sub_balance}@*)
    balances[to] += amount;
    ...
  }
  function _useAllowance(owner, spender, amount) {
    ...
    require(allowances[owner][spender] >= amount);(*@\label{code:allowance_check}@*)
    allowances[owner][spender] -= amount;
    ...
  }
}
\end{lstlisting}
\caption{A fragment of an ERC20 contract in Solidity~\cite{solidity}}
\label{fig:erc20}
\end{figure}


\noindent\textbf{An Example of Data Conflicts}.
\Cref{fig:erc20} presents a fragment of the implementation of an ERC20 contract.
The contract's persistent states are represented by two variables: 
the mapping \code{balances} that records the number of tokens owned by each account,
and the mapping \code{allowances} that tracks the number of tokens authorized for third-party transfers by owners.
Tokens can be transferred via function \code{transferFrom}, where message senders transfer a certain number of tokens on behalf of owners.
The function fails if the owner does not have enough tokens (line \ref{code:balance_check}) or the message sender does not have enough allowances (line \ref{code:allowance_check}).

In most cases, ERC20 contracts lead to data conflicts when multiple transactions distribute tokens from the same sender address~\cite{garamvolgyi2022utilizing}.
For example, consider the following two concurrent transactions: 
\begin{align*}
    tx_1 &= \text{\code{transferFrom}}_D(A,B,value_1) \\
    tx_2 &= \text{\code{transferFrom}}_E(A,C,value_2)
\end{align*}
\noindent,where the function subscript denotes the transaction sender address (i.e., the return value of \code{_msgSender()} in \Cref{fig:erc20}).
These two transactions conflict because they both read and write the same storage slot \code{balances[A]}.

\noindent\textbf{Transaction-level Conflict Handling Strategy.}
Traditional algorithms cannnot parallelize $tx_1$ and $tx_2$ due to the conflict on \code{balances[A]}.
Nevertheless, we find that most operations in both transactions do not depend on the conflict (i.e., \code{balances[A]}) and are thus conflict-free.
We refer to such conflict-handling strategies as \emph{transaction-level} strategies, because they view a transaction as a whole and are unaware of internal operations.
When encountering a conflict, they either block or abort the \emph{entire} transaction, which is inefficient under severe data contention.

\noindent\textbf{Operation-level Conflict Handling Strategy.}
Let us reconsider the previous example.
Under the optimistic concurrency control, $tx_2$ cannot pass the validation phase because it misses $tx_1$'s update on \code{balances[A]} and produces the wrong \code{balances[A]} (line \ref{code:sub_balance} of \Cref{fig:erc20}).
But note that $tx_2$ does correctly update all other storage slots like \code{balances[C]} and \code{allowances[A][E]}.
Inspired by the finding, we extend traditional OCC by introducing a \emph{redo phase} after the validation phase.
During the redo phase, rather than aborting $tx_2$ immediately after the validation fails, we re-execute line \ref{code:sub_balance} based on the correct \code{balances[A]} value.
In this way, we resolve the conflict on \code{balances[A]} by re-executing only one line of the source code and thus parallelize most operations in $tx_1$ and $tx_2$ successfully.
We refer to this conflict-handling strategy as the \emph{operation-level} strategy because it only re-executes operations that depend (directly and indirectly) on conflicts.
Considering that most conflicts in blockchain workloads only influence several operations (as shown in this ERC20 contract example), the operation-level strategy can significantly alleviate the performance degradation caused by conflicts.

To guarantee the correctness of the operation-level strategy, we further introduce a technique called \emph{constraint guards}.
We use another scenario from the previous example to illustrate the necessity of the constraint guards.
Suppose \code{balances[A]} does not own enough tokens to afford $tx_2$ after executing $tx_1$.
In this case, $tx_2$ should be aborted due to the balance checking in line \ref{code:balance_check}.
Our operation-level strategy should detect the violation and abort $tx_2$, instead of simply re-executing line \ref{code:sub_balance}.
To achieve this, when executing lines \ref{code:balance_check} and \ref{code:allowance_check}, we insert constraint guards that assert the required conditions must hold.
During the redo phase of $tx_2$, we re-check these constraints, detect that the correct \code{balances[A]} does not satisfy the constraint in line \ref{code:balance_check}, and thus abort $tx_2$.
\section{Challenges and Our Solution}
\subsection{Challenges}
Our main idea is to resolve transaction conflicts at the operation level.
We extend traditional OCC with a redo phase to identify and re-execute operations influenced by conflicts.
Nevertheless, both the conflicting operation identification and re-execution are challenging at the EVM bytecode level.

\noindent\textbf{Conflicting Operations Identification.}
To guarantee the correctness and efficiency of the redo phase, we must precisely identify all operations that depend on conflicts (i.e., without under-estimation or over-estimation).
However, it is difficult to figure out the dependencies at the EVM bytecode level, as EVM operations do not state the data dependencies explicitly.
In other words, they do not directly take the results of other operations as the inputs; instead, they retrieve inputs from the \emph{runtime context} (i.e., stack, memory, and storage).
This indirection makes it hard to identify the definition operations of these inputs.
    
\noindent\textbf{Conflicting Operations Re-execution.}
The execution of an EVM operation depends on its runtime context.
For example, to execute an \code{ADD} operation, we must know the top two elements in the stack; to execute a \code{MLOAD} operation, we must know the memory content at that point.
However, it is difficult to obtain the runtime context corresponding to a specific operation during the redo phase without executing all previous operations.
Without these run-time contexts, we cannot re-execute EVM operations correctly.


\subsection{Our Solution: SSA Operation Log}
\begin{figure}[t]
    \centering
    \includegraphics[width=0.39\textwidth]{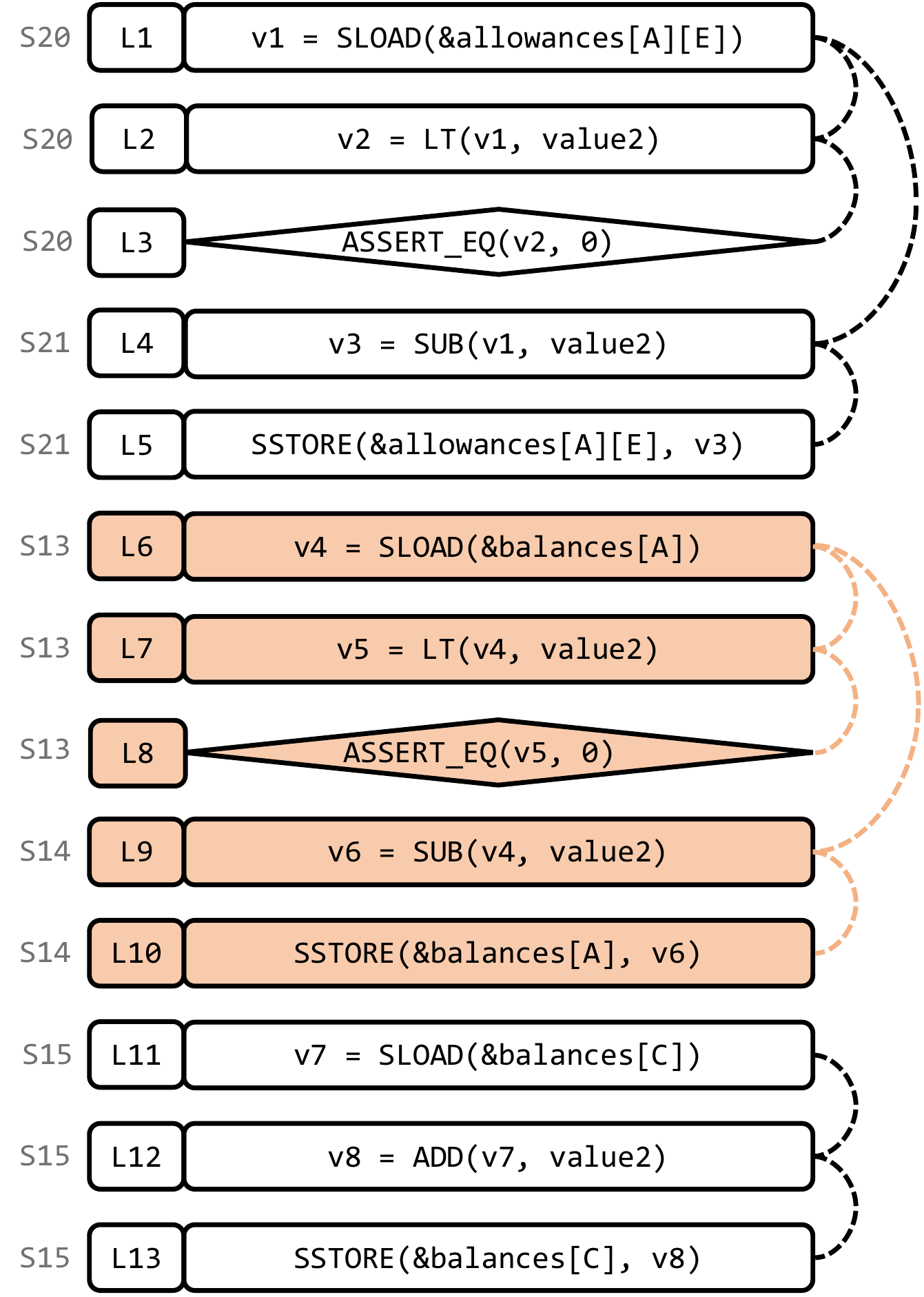}
    \caption{
    The SSA operation log of the transaction $tx_2$ in \Cref{sec:erc20}.
    Operations in diamond boxes (i.e., L3 and L8) are constraint guards.
    The corresponding line numbers in source code (\Cref{fig:erc20}) are annotated in the left; the definition-use and use-definition chains are drawn in the right.
    }
    \label{fig:oplog}
\end{figure}
\begin{figure*}
    \centering
    \includegraphics[width=0.85\textwidth]{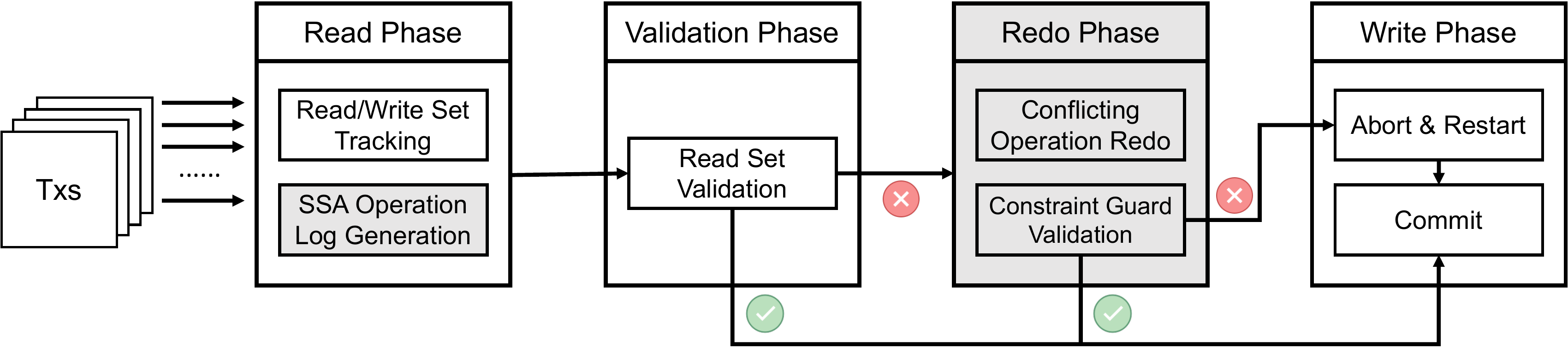}
    \caption{\name{} workflow. 
    }
    \label{fig:workflow}
\end{figure*}

To address these challenges, we propose a new technique called \emph{SSA operation log}, which records not only the operands and results of executed operations but also the data dependencies among these operations.
The main idea is that instead of analyzing EVM bytecode which depends on the runtime context, we dynamically generate the static single assignment form (SSA)~\cite{ssa} representation of executed operations.
More specifically, we enforce that the inputs of an operation must be explicitly represented by (i) immediate values, (ii) results of previous operations, and (iii) committed storage slots\footnote{We refer to a storage slot as committed if previous operations in the current transaction have not updated it.}.
In this way, we eliminate all runtime context accesses, making the conflicting operation identification and re-execution much easier.

The SSA operation log provides well-defined use-definition and definition-use\footnote{
When a variable, $v$, is on the left-hand side of an assignment statement, $s$, then $s$ is a definition of $v$.
When a variable, $v$, is on the right-hand side of a statement, $s$, then a definition of $v$ is used at $s$.
}
relationships among operations.
During the redo phase, we can identify all conflicting operations by traversing the definition-use chains, re-construct operation inputs based on their definition operations, and finally re-execute these conflicting operations.
For example, \Cref{fig:oplog} illustrates the SSA operation log of transaction $tx_2$ in \Cref{sec:erc20}.
All operations explicitly state their data dependencies.
During the redo phase, we can identify that the conflict on \code{balances[A]} is introduced by the \code{L6} entry.
Following the definition-use chains depicted at the right of \Cref{fig:oplog}, we then find that operations from \code{L6} to \code{L10} are influenced by \code{balances[A]}.
Therefore, we only need to re-execute these five operations one by one.
\section{Design and Implementation}
In this section, we elaborate on our operation-level concurrency control algorithm designed for blockchain workloads.
We have implemented an open-source prototype, \name{}, based on Go Ethereum (aka Geth)~\cite{geth}.
It is worth noting that the design rationale presented in this paper can also be applied to other blockchain systems.
Although the implementation may vary, our prototype provides detailed guidance for resolving transaction conflicts at the operation level and exploiting operation-level parallelism.

\subsection{Overview}
As shown in \Cref{fig:workflow}, \name{} uses a variant of optimistic concurrency control algorithms.
However, \name{} does not abort transactions immediately after the validation phase fails.
Instead, it introduces a \emph{redo phase} to resolve conflicts at the operation level.
More specifically, \name{} executes transactions in the following four phases:

\begin{enumerate}
    \item \emph{Read phase,} where \name{} executes transactions concurrently and speculatively.
    As the traditional OCC, \name{} tracks all key-value pairs a transaction reads and writes.
    In addition, \name{} dynamically generates an \emph{SSA operation log} for each transaction (\Cref{sec:oplog}).
    
    \item \emph{Validation phase}, where \name{} validates the correctness of the speculative execution in the read phase.
    A transaction enters the validation only if all previous transactions commit.
    During the validation phase, \name{} re-reads all key-value pairs in the transaction's read set from the committed storage and examines whether values read in the previous phase are consistent with the newly read values.
    If the validation succeeds, the transaction enters the write phase; otherwise, it enters the redo phase.
    
    \item \emph{Redo phase}, where \name{} attempts to resolve conflicts by re-executing all conflicting operations.
    Leveraging the operation dependencies record in the SSA operation log, \name{} validates constraint guards and identifies and re-executes conflicting operations (\Cref{sec:redo}).
    If all constraints are satisfied, the transaction enters the write phase and commits; otherwise, the transaction has to be aborted and restarted in the write phase.
    
    \item \emph{Write phase}, where \name{} finalizes transactions.
    ParallelEVM directly commits the transaction by installing its write set to the storage if a transaction passes either the validation phase or the redo phase.
    Otherwise, \name{} has to abort, restart and finally commit the transaction.
\end{enumerate}


\subsection{SSA Operation Log Generation}\label{sec:oplog}
This section shows how to dynamically generate an SSA operation log in the read phase.
An entry in the SSA operation log typically consists of the following fields:
\begin{itemize}
    \item \code{LSN}, a unique log sequence number that identifies the entry.
    \item \code{Opcode}, the operation code of the operation.
    \item \code{Operands}, the inputs of the operation.
    \item \code{Result}, the output of the operation.
    \item \code{Def}, the definition operations of the inputs. 
    This field comprises three sub-fields, namely \code{stack}, \code{storage}, and \code{memory}, which track the data dependencies via the stack, storage, and memory.
    We elaborate on the detailed structures and generation rules for them in Sections \ref{sec:stack}-\ref{sec:memory}, respectively.
\end{itemize}

The fundamental principle behind the generation algorithm is to track the data flow and figure out the definition operations of instruction inputs.

\subsubsection{Stack Operation Log Generation}\label{sec:stack}
First, we present how to generate SSA operation log entries for operations that access only the stack.
For each operation, \name{} records the definition operations of all stack operands in the \code{def.stack} field.
That is, if a stack operand, \code{operands[i]}, is the result of a previous operation, the \code{def.stack[i]} field is set to the \code{LSN} of that operation; otherwise, the \code{def.stack[i]} field is set to \code{NULL}.

To generate the \code{def.stack} field correctly, \name{} employs a technique called \emph{shadow stack}, which records the definition operation of each stack item.
If a stack item is the \code{result} of a previous log entry, \name{} sets the corresponding shadow stack item to the \code{LSN} of that entry; otherwise, \name{} sets the shadow stack item to \code{NULL}.
In this way, for each operation, \name{} can obtain the definition operations of its stack operands from the shadow stack and store them in the \code{def.stack} field.

\name{} manipulates the shadow stack in a similar way as the stack.
For the \code{PUSH} operation, which pushes a constant data to the stack, \name{} pushes \code{NULL} to the shadow stack.
For the \code{POP}, \code{SWAP} or \code{DUP} operation, \name{} pops, swaps or duplicates the shadow stack.
For other computational operations, \name{} first obtains and pops all shadow stack items (denoted as \code{lsns}) corresponding to the stack operands.
If \code{lsns} are all \code{NULL}s, which indicates that all operands are constants and thus the result is also a constant, \name{} pushes \code{NULL} to the shadow stack.
Otherwise, \name{} generates a new log entry, stores the \code{lsns} in the \code{def.stack} field, and pushes the \code{LSN} of the entry to the shadow stack.

\begin{figure}
    \centering    \includegraphics[width=0.45\textwidth]{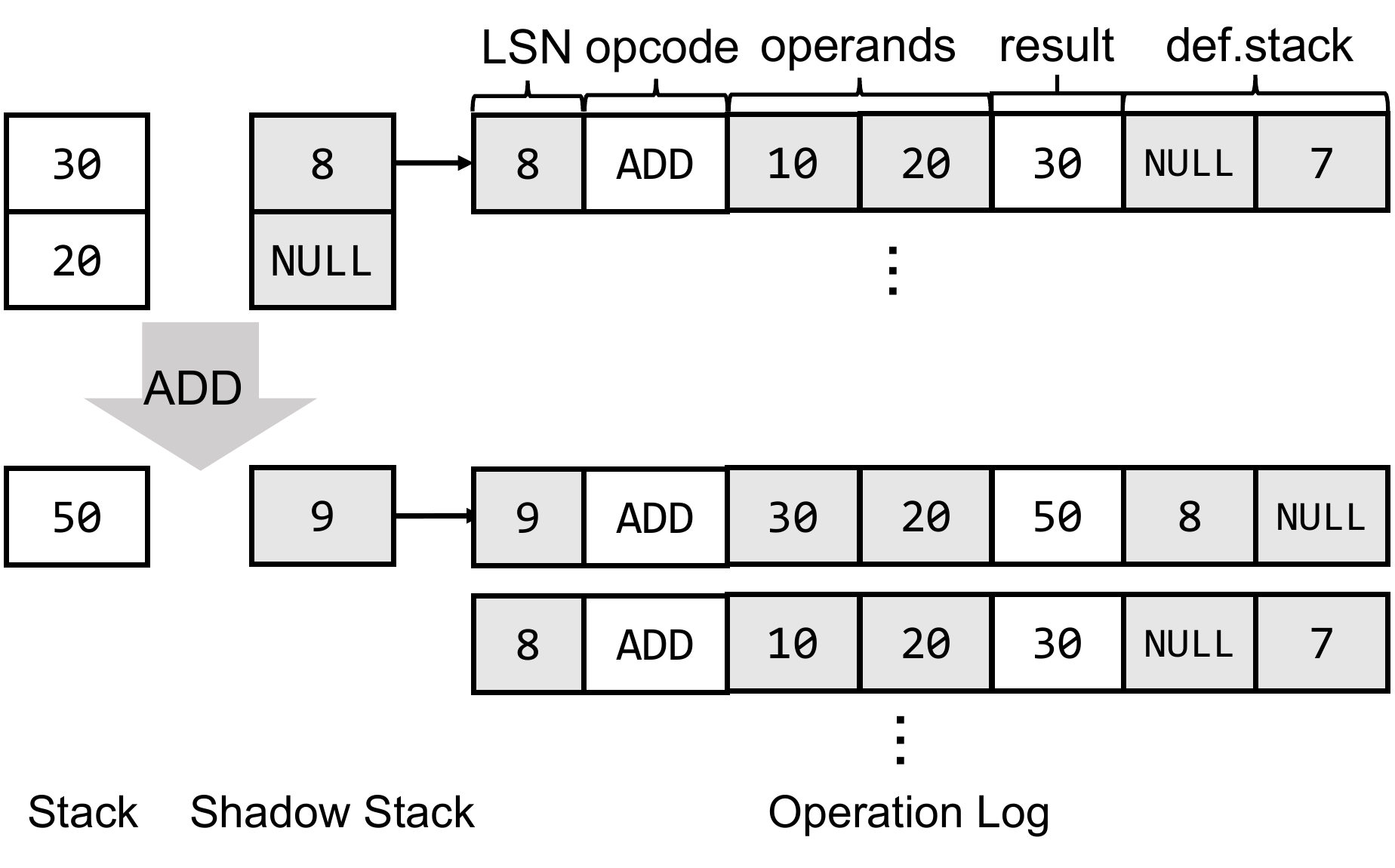}
    \caption{Stack tracking example}
    \label{fig:shadow_stack}
\end{figure}

\Cref{fig:shadow_stack} shows an example of generating a log entry for an \code{ADD} operation.
Before executing the \code{ADD}, the stack contains two items (i.e., \code{30} and \code{20}).
According to the shadow stack, we can infer that the first item (i.e., \code{30}) is the result of the eighth log entry, while the second item (i.e., \code{20}) is a constant value.
After the execution, \name{} generates a new log entry for the \code{ADD} operation (i.e., the ninth entry), pushes the computation result (i.e., \code{50}) to the stack, and finally pushes the \code{LSN} of the new entry (i.e., \code{9}) to the shadow stack.
In this way, the following operations can realize that the top stack item, \code{50}, is the result of the ninth log entry.

For storage and memory operations, the generation rules for \code{def.stack} fields are similar.
But they should additionally maintain the \code{def.storage} and \code{def.memory} fields.

\subsubsection{Storage Operation Log Generation}\label{sec:storage}
Then we consider storage operations, namely \code{SLOAD} and \code{SSTORE}.
There are two types of \code{SLOAD}s: (i) reading committed storage and (ii) reading values written by previous \code{SSTORE}s in the current transaction.
\name{} uses the \code{def.storage} field to distinguish them.
For the first case, the \code{def.storage} field should be \code{NULL}, because such \code{SLOAD}s do not depend on any previous operation; while for the second case, the \code{def.storage} field should be the \code{LSN} of the latest \code{SSTORE} entry that writes to the storage slot.

\name{} maintains two maps to track the data flow over the storage.
The first map, \code{latest_writes}, records the \code{LSN} of the latest \code{SSTORE} on each storage slot.
The second map, \code{first_reads}, records the \code{LSN}s of \code{SLOAD}s that read each committed storage slot directly.
\name{} uses the \code{lastest_writes} map to generate the SSA log entry for storage operations, while uses the \code{first_reads} map to identify conflicting storage operations.
When generating a log entry for the \code{SSTORE} operation that writes to the storage slot $A$, \name{} updates the \code{latest_writes[A]} to the entry's \code{LSN}.
When generating a log entry for the \code{SLOAD} operation that reads from storage slot $A$, if $A$ does not exist in the \code{latest_writes} map, \name{} sets the \code{def.storage} field to \code{NULL} and appends the entry's \code{LSN} to the \code{first_reads[A]}; otherwise, \name{} sets the \code{def.storage} field to the value of \code{latest_writes[A]}.

\subsubsection{Memory Operation Log Generation}\label{sec:memory}
\begin{figure}
    \subfloat[Interleaved \code{MSTORE} and \code{MSTORE8}]{
        \includegraphics[width=0.45\textwidth]{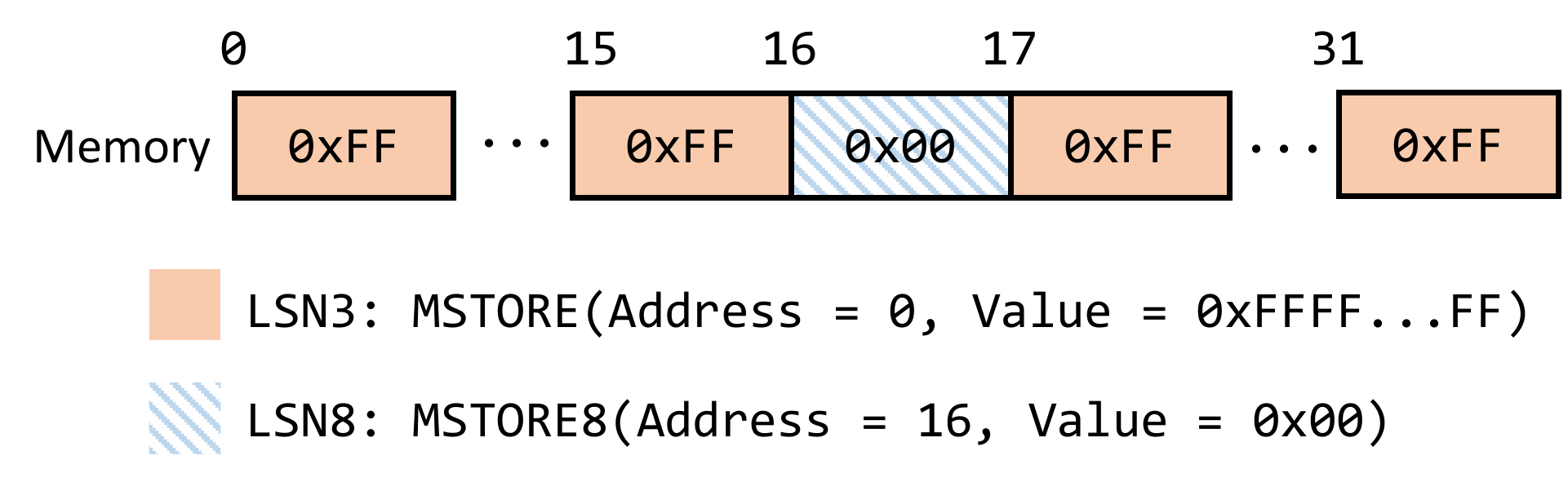}
        \label{fig:interlevae_mstore}
    }
    
    
    \subfloat[Shadow memory]{
    \includegraphics[width=0.45\textwidth]{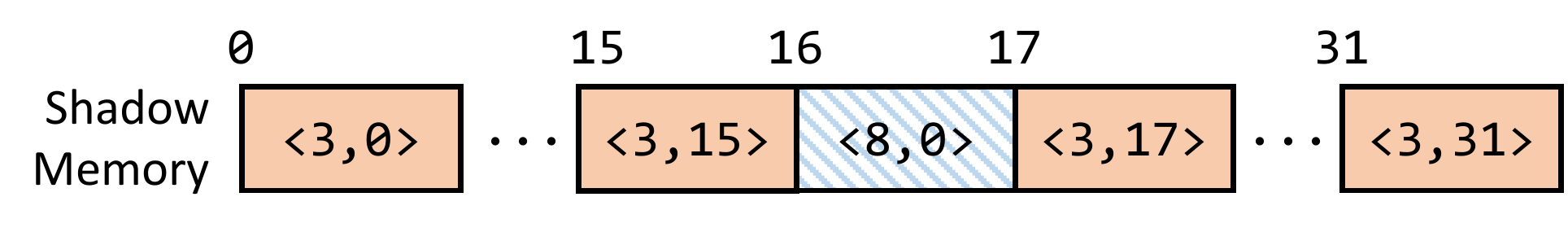}
    \label{fig:shadow_memory}
    }

    
    \subfloat[The \code{def.memory} field corresponding to (b)]{
    \includegraphics[width=0.43\textwidth]{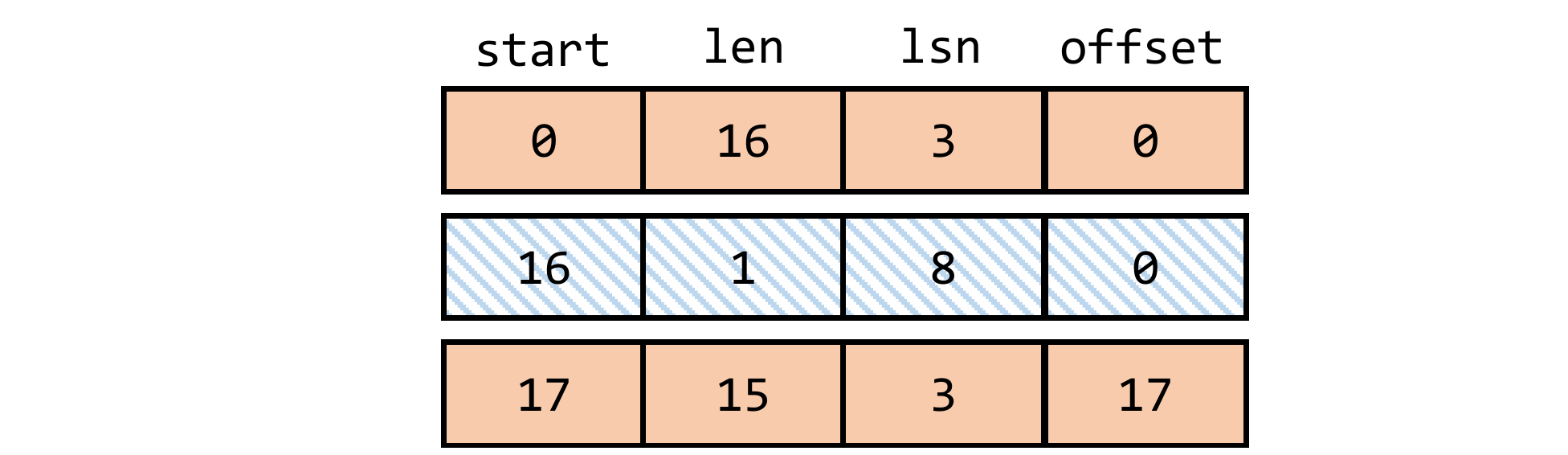}
    \label{fig:prev_mstore}
    }
    
    \caption{Memory tracking example}
    \label{fig:memory}
\end{figure}

In EVM, memory operations (e.g., \code{MLOAD} and \code{MSTORE*}) are not aligned.
Therefore, multiple memory writes may overlap, and a single memory read may depend on multiple memory writes.
\Cref{fig:interlevae_mstore} shows an example of interleaved \code{MSTORE} and \code{MSTORE8}.
In this case, the following memory operations that read these bytes must be aware that they depend on these two \code{MSTORE*} operations.

To track the data flow over the memory, \name{} employs \emph{shadow memory}, a technique similar to the shadow stack.
As shown in \Cref{fig:shadow_memory}, for every byte set by a memory write, \name{} set the corresponding data in the shadow stack to \code{<LSN, offset>}, where \code{LSN} denotes the log sequence number of that write operation and \code{offset} denotes the distance between the current byte and the first byte set by the memory write operation.

Moreover, \name{} uses the \code{def.memory} field to record the data dependencies for every memory read operation based on the shadow memory.
Specifically, the \code{def.memory} field consists of multiple \code{<start, len, lsn, offset>} tuples, which denotes that the \code{[start:start+len)} bytes in the memory input are set by the \code{[offset:offset+len)} bytes in the \code{result} of the \code{lsn}-th log entry.
For instance, if a \code{MLOAD} operation reads the 32-byte memory region in \Cref{fig:interlevae_mstore}, \name{} will set the \code{def.memory} field to the tuples shown in \Cref{fig:prev_mstore}.

\subsubsection{Constraint Guard Generation}
In order to guarantee correctness, \name{} must satisfy certain constraints in the redo phase.
First, the control-flow constraints state that the re-execution in the redo phase must follow the same path as the speculative execution in the read phase.
For example, for a conditional jump (e.g., \code{JUMPI}), the jump condition must remain the same.
Similarly, for an unconditional jump (e.g., \code{JUMP}), the jump destination must not change.
Therefore, when executing control-flow-related operations, \name{} inserts constraints guards to ensure that control-flow-related operands (e.g., jump conditions and jump destinations) are unchanged.
More specifically, if the shadow stack item corresponding to a control-flow-related operand is not \code{NULL} (i.e., the operand is not constant), \name{} inserts a special \code{ASSERT_EQ} log entry whose \code{operands} field is the operand value and \code{def.stack} field is the shadow stack item.
In this way, during the redo phase, \name{} can compare the \code{operands} field and the \code{result} field of the log entry whose \code{LSN} is \code{def.stack}.
If these two fields match, the constraint is satisfied; otherwise, the redo phase should fail.

Second, \name{} also requires data flow constraints to ensure that the re-execution does not alter the data dependencies among operations.
For example, for a \code{MSTORE} operation, \name{} must ensure that its target address is unchanged in the redo phase.
Otherwise, \name{} cannot figure out which operation depends on the address-changing \code{MSTORE}.
To achieve this, if the target address of a run-time context access operation (e.g., \code{MLOAD} and \code{MSTORE}) is not constant, \name{} inserts a \code{ASSERT_EQ} log entry to guard the address operand.

\subsubsection{Definition-Use Graph Generation}
Based on the \code{def} fields, \name{} can find all operations that a specific operation depends on.
Nevertheless, the conflicting operation identification in the redo phase requires a reverse procedure --- find all operations depending on a specific operation.
To achieve this, \name{} constructs a definition-use graph, $DUG$, while generating the SSA operation log.
Each log entry becomes a node in the graph, and each edge from $entry1$ to $entry2$ denotes that $entry2$ uses the result of $entry1$.


\begin{algorithm}[t]
\caption{The algorithm for the redo phase}
\label{algor:redo}
\small
\SetKw{forin}{in}
\SetKw{is}{is}
\SetKw{isnot}{is not}
\SetKw{all}{all}
\SetKw{and}{and}
\SetKw{false}{false}
\SetKw{true}{true}
\SetKw{continue}{continue}

\KwIn{SSA operation log: $oplog$\; 
definition-use graph: $DUG$\;
conflicting storage slot map: $conflicts[keys]values$.
}
\KwOut{whether the redo phase is successful.}
\SetKwFunction{FMain}{Redo}
\SetKwProg{Fn}{Function}{:}{end}
\Fn{\FMain{$oplog$, $DUG$, $conflicts$}}{
    $sources \leftarrow$ \code{SLOAD}s in $oplog$ that read $conflicts$ directly\;\nllabel{algor:get_source}
    \For{entry \forin{} sources}{\nllabel{algor:update_sload_beg}
        $entry.result \leftarrow conflicts[entry.operands]$\;
    }
    \nllabel{algor:update_sload_end}
    $conflicting\_ops \leftarrow$ DFS($DUG$, $sources$)\;\nllabel{algor:def}
    \For{$entry$ \forin{} $conflicting\_ops$ / $sources$} {
        \uIf{entry.opcode \is{} \code{ASSERT_EQ}}{
            $expect \leftarrow entry.operands.value$\;\nllabel{algor:guard_beg}
            $current \leftarrow oplog[entry.operands.lsn].result$\;
            \lIf{$current \neq expect$}{\KwRet{false}}\nllabel{algor:guard_end}
        }
        \Else {
            reconstruct the inputs based on $entry.def$\;\nllabel{algor:reconstruct}
            re-execute $entry$ and update $entry.result$\;\nllabel{algor:reexec}
        }
    }
    \KwRet{true}
}
\end{algorithm}

\subsection{Redo Phase}\label{sec:redo}
\Cref{algor:redo} describes how we identify and re-execute conflicting operations in the redo phase.
When a transaction fails the validation phase and enters the redo phase, \name{} can obtain all conflicting storage slots and their correct values (i.e., the \code{conflicts} map in \Cref{algor:redo}).
During the redo phase, \name{} uses the \code{first_reads} map (introduced in \Cref{sec:storage}) to locate all conflicting \code{SLOAD} log entries that read the committed but conflicting storage slots (line \ref{algor:get_source}) and updates their results to the correct values (lines \ref{algor:update_sload_beg}-\ref{algor:update_sload_end}).
\name{} then applies the depth-first search algorithm to the definition-use graph $DUG$ (line \ref{algor:def}).
The search starts at the conflicting \code{SLOAD} entries, traverses all connected entries, and returns all conflicting operations.
After that, \name{} re-executes these conflicting operations one by one (except the previous \code{SLOAD} entries that have already been re-executed).
If an entry is a constraint guard, \name{} examines whether the constraint is still satisfied (lines \ref{algor:guard_beg}-\ref{algor:guard_end}).
Otherwise, \name{} re-constructs the operation inputs based on the results of all definition operations listed in the \code{def} field (line \ref{algor:reconstruct}) and then re-executes the operation (line \ref{algor:reexec}).

\subsection{Implementation Details}
We have implemented \name{} in Go Ethereum (version 1.10.17).
It consists of about 4200 lines of code changed: about 2600 for the optimistic concurrency control framework, and about 1600 for the operation-level conflict-handling strategy.
Besides the design highlights presented above, \name{} also resolves the balance and nonce conflicts at the operation level using similar techniques.
Moreover, we implement \name{} carefully and take into account corner cases of Ethereum, such as dynamic gas instructions and implicit gas limit checking.
\begin{table}[t]
    \centering
    \footnotesize
    \begin{tabular}{c @{\hspace{1.25\tabcolsep}} c @{\hspace{1.25\tabcolsep}} c @{\hspace{1.25\tabcolsep}} c @{\hspace{1.25\tabcolsep}} c @{\hspace{1.25\tabcolsep}} c}
    \toprule
    Baseline & OCC & \name{} & Prefetch & OCC+ & \name{}+\\
    \midrule
    1$\times$ & 2.49$\times$ & \textbf{4.28$\times$} & 2.89 $\times$ & 3.25$\times$ & \textbf{7.11$\times$}\\
    \bottomrule
\end{tabular}
    \caption{Speedups achieved by different strategies. OCC+ denotes OCC with prefetching, and \name{}+ denotes \name{} with prefetching.}
    \label{tab:speedup}
\end{table}

\begin{figure}[t]
    \includegraphics[width=0.47\textwidth]{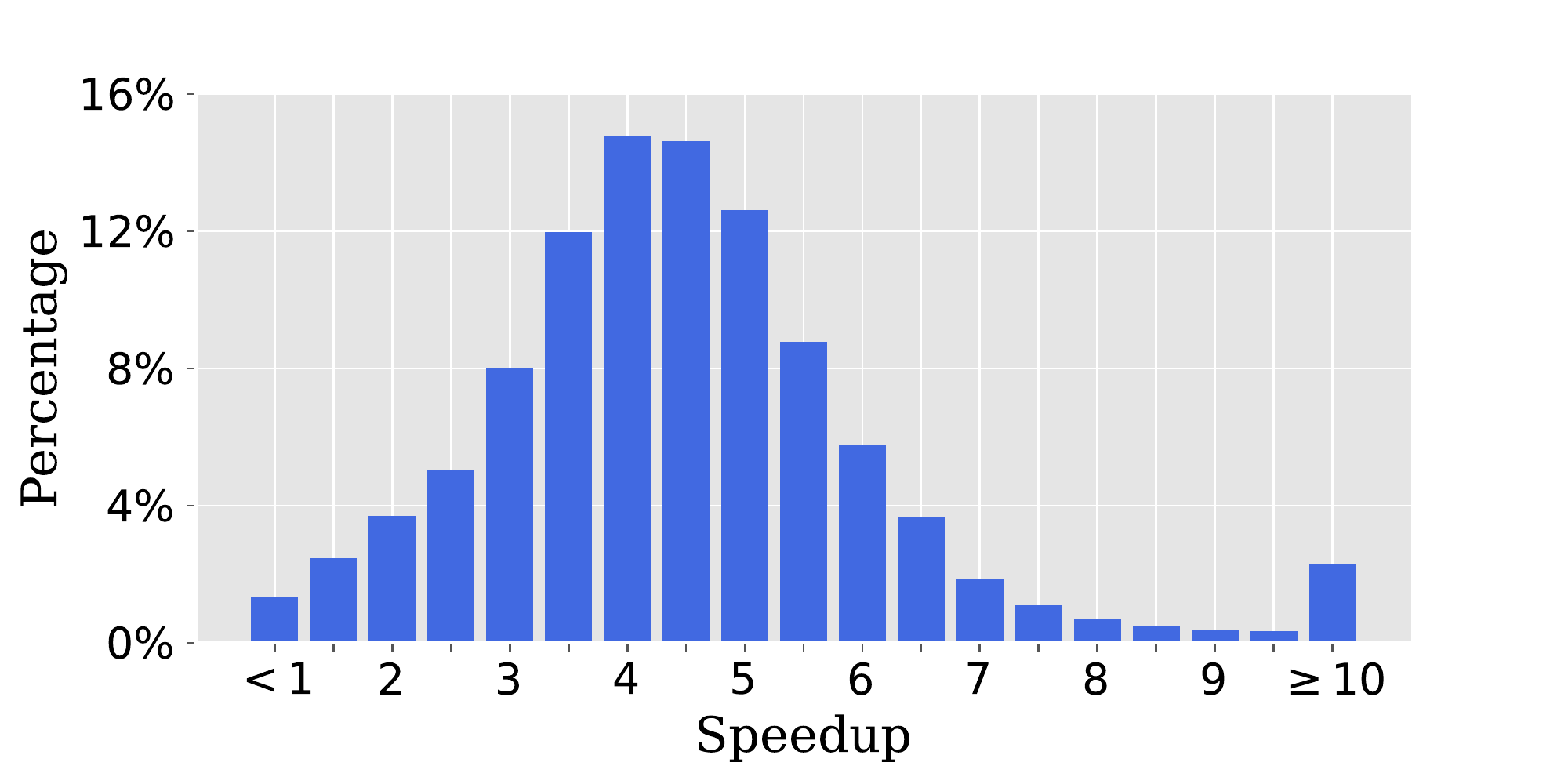}
    \caption{Speedup distribution for \name{}}
    \label{fig:speedup}
    
    \includegraphics[width=0.47\textwidth]{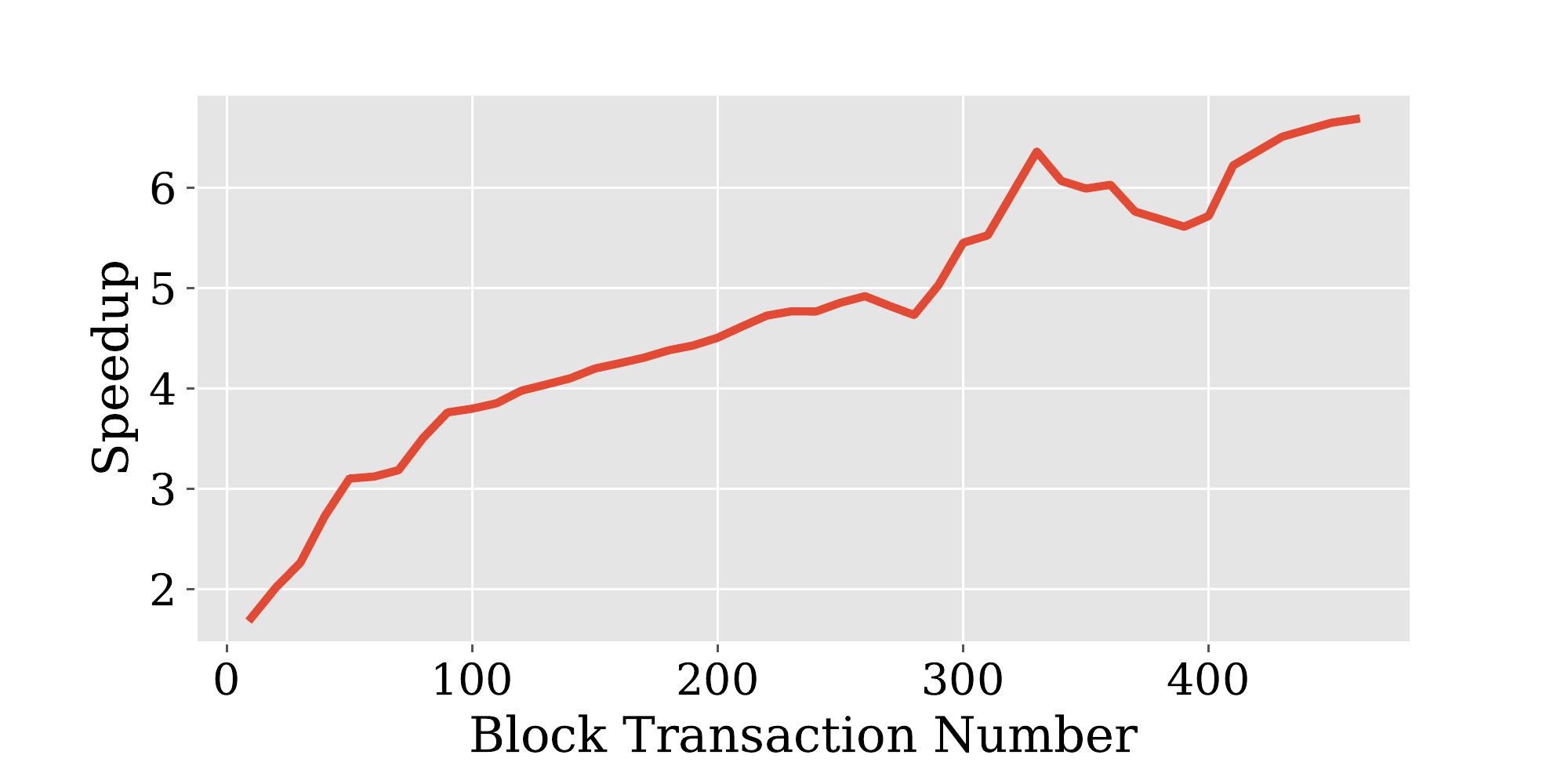}
    \caption{The correlation between block transaction number and average speedup}
    \label{fig:speedup_txn}
\end{figure}

\section{Evaluation}
In this section, we present a comprehensive evaluation of \name{} on its correctness and performance.
Our experiments are conducted on a platform with an Intel i7-10700F 2.9GHz processor (8 physical cores and 16 virtual cores), 16GB memory, and a 1TB SSD.
The system software running on the platform is Ubuntu 22.04 LTS with Linux kernel 5.15 and Ext4 file system.

\subsection{Correctness Validation}
The correctness of \name{} can be validated by the design of Ethereum itself.
Ethereum maintains its state as a Merkle Patricia Trie (aka MPT)~\cite{mpt}, where every non-leaf node contains the cryptographic hash of its child nodes.
Therefore, two Ethereum states are identical if and only if the root nodes of their MPTs are identical.
Moreover, for validation purposes, every Ethereum block contains the cryptographic hash of the MPT root node after all transactions in the block are correctly executed.
We have run \name{} to process the initial 11.5 million blocks from the Ethereum mainnet, and \name{} always produced the MPT root nodes that matched the values in the block.
The result demonstrates that \name{} rigorously follows the rules defined in the Ethereum yellow paper~\cite{wood2014ethereum}.

\subsection{Performance Evaluation}
\noindent\textbf{Speedup.}
To measure the performance of \name{}, we run \name{} to process 500k real-world Ethereum blocks (the block numbers range from 11 million to 11.5 million).
\name{} achieves an average of 4.28$\times$ speedup compared to Geth 1.10.17 (baseline).
We show the distribution of speedups in \Cref{fig:speedup}: most blocks are accelerated by 2\textasciitilde 7$\times$, only 0.88\% are not accelerated, and 2.66\% are accelerated by more than 10$\times$.
We also compare \name{} with the traditional OCC.
As shown in \Cref{tab:speedup}, OCC obtains only 2.49$\times$ speedup in real-world Ethereum workloads, which demonstrates that \name{} significantly outperforms traditional concurrency control algorithms.

Besides, \Cref{fig:speedup_txn} illustrates the correlation between the block transaction number and the speedup achieved by \name{}.
\name{} generally obtains higher speedups on blocks that contain more transactions (i.e., with larger block sizes).
It indicates that if the block sizes are enlarged in the future, which is the goal of transaction execution acceleration, \name{} could achieve a higher speedup.

\begin{figure}
    \centering
    \includegraphics[width=0.47\textwidth]{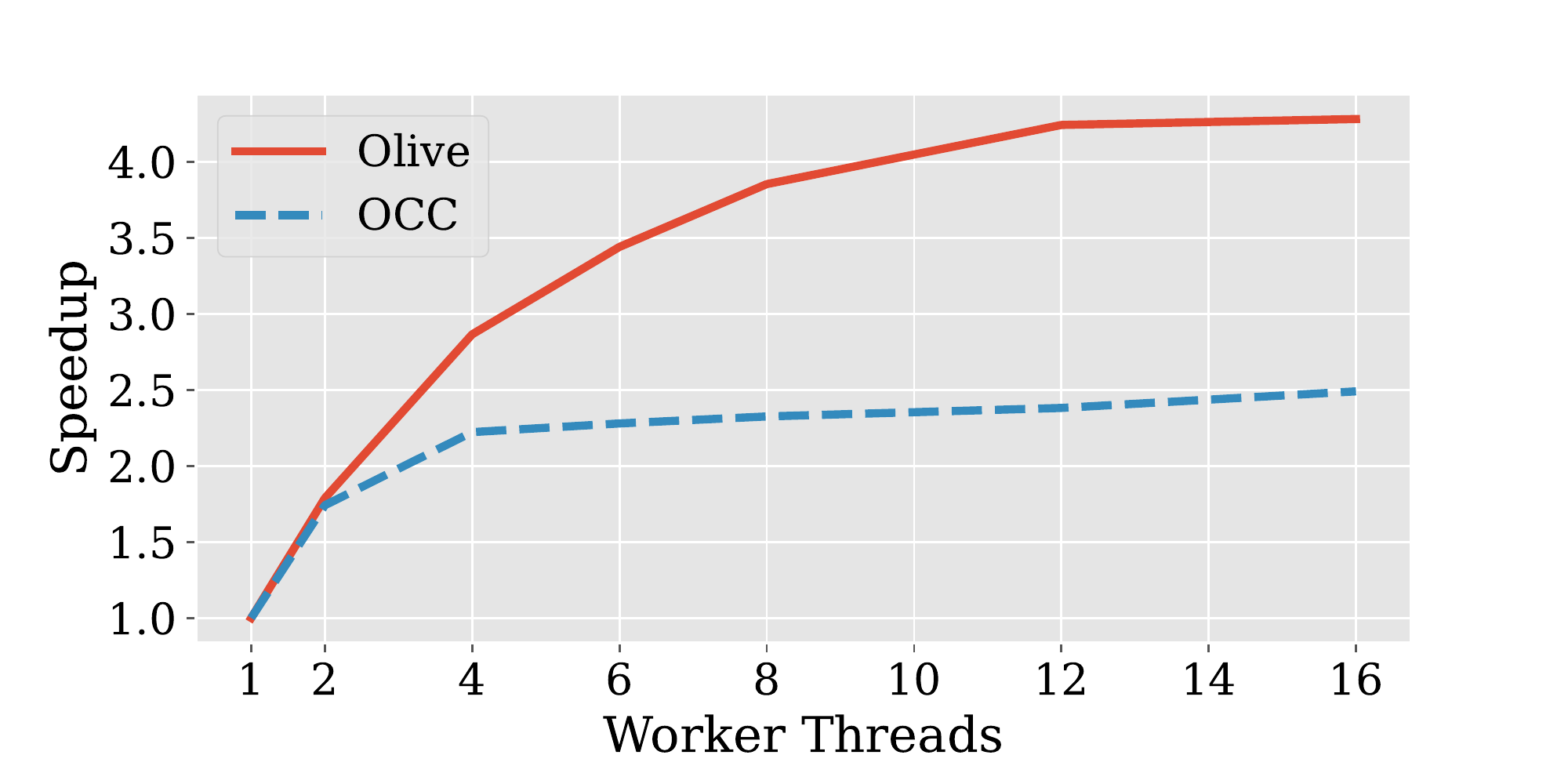}
    \caption{Speedups of \name{} and OCC}
    \label{fig:scalable}
\end{figure}
\noindent\textbf{Scalability.}
\Cref{fig:scalable} shows the speedups of \name{} and OCC as we increase the number of worker threads.
\name{} scales well to eight threads, while OCC only scales to four threads.
The speedups of \name{} increase slightly past eight threads, because there are only eight physical cores on our platform.
More than eight worker threads would contend for cores, even the hyperthreading is enabled.

\begin{figure}
    \centering
    \includegraphics[width=0.47\textwidth]{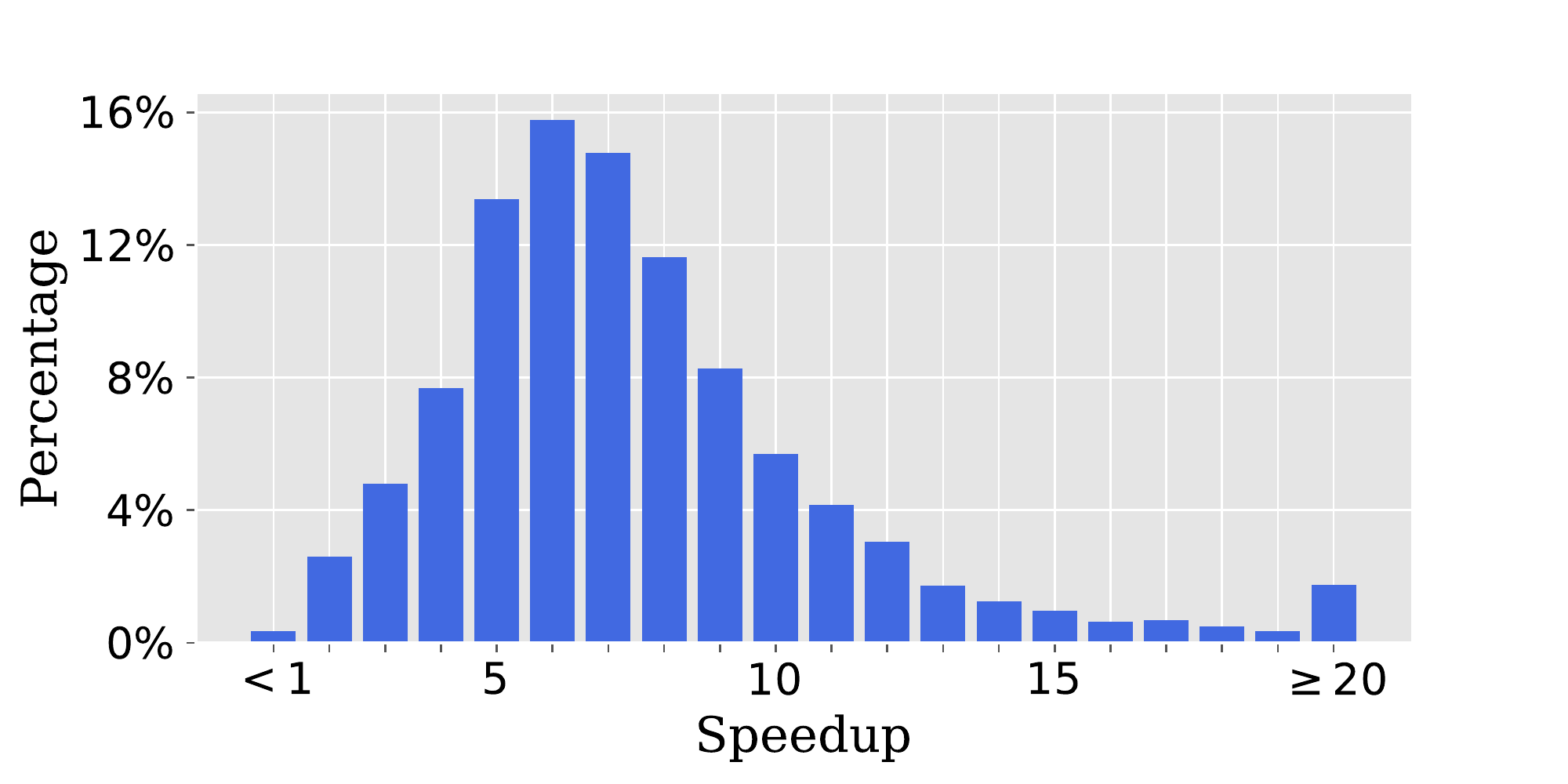}
    \caption{Speedup distribution for \name{}+}
    \label{fig:prefetch_speedup}
\end{figure}

\noindent\textbf{Speedup with Prefetching.}
Analyzing the runtime profiling data, we identify that the performance bottleneck of \name{} is storage operations, especially \code{SLOAD}s.
Accessing the persistent Ethereum state involves loading values from the on-disk LevelDB database, which incurs high latency.
Previous works have proposed a state prefetching technique to address the problem~\cite{chen2021forerunner}.
On Ethereum, transactions are known to most participating nodes before they are packed into blocks by the miners.
Therefore, to avoid the expensive disk read in the actual execution, one can pre-execute a pending transaction to predict which storage slots it may access and then prefetch and cache them in the memory.
To evaluate how many speedups \name{} can achieve if combined with the state prefetching technique, we modify the import functionality of Geth so that every block is processed twice.
Specifically, we prefetch all required storage slots to the memory in the first execution and measure the performance of the second execution.
As shown in \Cref{tab:speedup}, with the help of the prefetching technique, \name{} further achieves an average speedup of 7.11$\times$.
We report the speedup distribution in \Cref{fig:prefetch_speedup}.
As comparisons, the prefetching technique itself and OCC with prefetching accelerate the execution by 2.89$\times$ and 3.25$\times$, respectively.
These results suggest that \name{} can cooperate better with the prefetching technique than traditional algorithms.


\noindent\textbf{SSA Operation Log.}
To further illustrate the effectiveness of our operation-level conflict-handling strategy, we conduct detailed measurements on the SSA operation log generation and re-execution.
Experiment results show that our SSA operation log generation algorithm only incurs an average of 4.5\% runtime performance overhead for each transaction.
Besides, our generation algorithm reduces the SSA operation log size significantly.
The average EVM instruction count of contract invocation is 2559, while the average SSA operation log length is 127.
In other words, the size of SSA operation log, including additional constraint guards, is only 5.0\% of the original EVM instructions.
This is mainly because our generation algorithm reduces the stack manipulation instructions (e.g., \code{PUSH} and \code{POP}) and instructions that do not depend on storage slots.
Moreover, only seven log entries (0.3\% of the original EVM instructions) on average are re-executed in the redo phase, and the time spent on the redo phase is only 4.9\% of the overall block processing time.
These results demonstrate the efficiency of the redo phase.

\noindent\textbf{Resource Utilization.}
Finally, we report the resource utilization of current implementation of \name{} compared to the official Geth (v1.10.17).
On our platform, the average memory consumption of \name{} is 9.48 GB, and the average utilization of 16 vCPUs is 32.19\%.
In contrast, the official Geth consumes 9.08GB memory and utilize 19.38\% of all CPU resources.
In other words, \name{} incurs only 4.41\% memory overhead (caused by the shadow stack and shadow memory) while increases the overall CPU utilization by 66.10\% (thanks to the parallel transaction execution).
The overall CPU utilization of \name{} is not increased dramatically, because \name{} only parallelize the transaction execution component of Geth.
Many other components, like the expensive state committing, are still single-threaded.

\section{Related Work}
Existing works for improving the throughput of blockchains can be broadly categorized into two classes.
The first class aims to enlarge the block size by accelerating the transaction execution, while the second class seeks to shorten the block time by enabling faster block generation rates.

\noindent\textbf{Transaction Execution Acceleration.}
The concurrent execution of blockchain transactions has become the focus of much research in recent years.
Saraph et al.~\cite{saraph} proposed a parallel execution engine, where they execute all transactions in parallel, roll back transactions that cause data conflicts and then re-execute these conflicting transactions sequentially.
They also observed that several popular contracts caused severe contention and performance degradation.
Garamvölgyi et al.~\cite{garamvolgyi2022utilizing} conducted an empirical study on the parallelism of the Ethereum workloads and identified that most data conflicts derive from counters (e.g., balances).
They further introduced partitioned counters and special commutative instructions to alleviate data conflicts.
Dickerson et al.~\cite{dickerson} described a parallel smart contract execution approach for miners and validators based on techniques adapted from software transactional memory~\cite{HybridTransactionalMemory,herlihy2008transactional,Herlihy2003SoftwareTM}.
They allow miners to execute smart contracts in parallel and resolve data conflicts by delaying or rolling back some conflicting invocations.
Besides, miners also generate concurrent schedules dynamically so that validators can convert the schedules into deterministic, parallel fork-join programs.
Similarly, Anjana et al.~\cite{optsmart} harness optimistic software transactional memory systems to execute transactions concurrently in miners.
They also introduce block graphs to capture the dependency of transactions in miners and allow validators to re-execute transactions concurrently and deterministically based on the graphs.
Zhang et al.~\cite{mvto} employed the multi-version timestamp ordering (MVTO) scheme to reduce transaction aborts in the optimistic concurrency control.
Jin et al.~\cite{cc-permissoned} proposed a concurrency protocol that focuses on the validator-side concurrency in permissoned blockchains.
They divide the transaction dependency graph generated by miners into several sub-graphs to preserve parallelism and reduce communication costs.
They further integrated their concurrency control algorithm with the PBFT consensus protocol.
Nevertheless, all these works use traditional transaction-level conflict-handling strategies.
Therefore, in real-world blockchain workloads, they encounter frequent delays or aborts and thus suffer severe performance degradation.
\name{}, instead, uses an operation-level strategy where most operations, even in conflicting transactions, can be executed
concurrently without blocking or aborting.

Speculative execution is another approach to speed up transaction execution~\cite{chen2021forerunner,JonathanEfficiently,Zyzzyva,Nightingale2005SpeculativeEI,MichaelPerformance}.
It utilizes the time window between when a transaction is known and when it is executed to predict and pre-execute transactions speculatively.
Forerunner~\cite{chen2021forerunner} proposed a constraint-based approach for speculative pre-execution.
It speculates on multiple futures and accelerates transactions based on imperfect predictions whenever certain constraints are satisfied.
It is worth noting that Forerunner is complementary to \name{}: Forerunner focuses on the pre-execution phase, while \name{} concentrates on the execution phase.
We have shown that \name{} can cooperate well with the state prefetching techniques.
Therefore, we believe that \name{} can achieve even better speedup if combined with more sophisticated speculative execution approaches such as Forerunner.

\noindent\textbf{Block Generation Acceleration.}
To reduce the consensus time between consecutive blocks, researchers have proposed various consensus algorithms based on Byzantine fault tolerance (BFT) protocols~\cite{Algorand,ByzCoin,pass2016hybrid,mazieres2015stellar,miller2016honey}.
Besides, to replace the longest chain rule used by PoW, researcher have explored alternative structures to organize blocks, such as the DAG-based chains~\cite{lewenberg2015inclusive,phantom,sompolinsky2015secure,conflux,sompolinsky2016spectre}, hierarchical chains~\cite{FruitChains,eyal2016bitcoin} and parallel chains~\cite{yu2020ohie}.
Although these techniques achieve higher block generate rates, most of them suffer scalability issues or live attacks~\cite{li2020survey}.
In contrast, \name{} can accelerate the transaction execution without sacrificing security and have full backward compatibility.
Moreover, \name{} is orthogonal to these techniques.
Therefore, the overall throughout of blockchains can be multiplied if \name{} is combined with these approaches properly.

Sharding is a popular solution to address the scalability issue of blockchains~\cite{cosplit,ELASTICO,elrondhighly,monoxide,OmniLedger,vault,al2017chainspace,HungTowards,RapidChain,zilliqa}.
Sharding systems split the blockchain states into multiple shards and select a small committee to process transactions on each shard.
Nevertheless, they also suffer security issues due to the slow committee reconfiguration.
Moreover, sharding exacerbates the hot spot problem in blockchain workloads: most popular contract invocations are likely to be processed in several hot shards, and transactions on each shard are more likely to conflict.
\name{} can help sharding systems resolve these conflicts at the operation level and mitigate the hot spot problem.

\section{Conclusion}
In this work, we present, implement and evaluate \name{}, an operation-level concurrent transaction execution system for Ethereum.
We resolve data conflicts by identifying and re-executing conflicting operations, so that all other conflict-free operations can be executed concurrently without blocking or aborting.
Evaluation results demonstrate that \name{} can effectively accelerate transaction execution in real-world Ethereum workloads.

\bibliographystyle{plain}
\bibliography{references}

\end{document}